# Model and Visualization of Ray Tracing using JavaScript and HTML5 for TIR Measurement System Equipped with Equilateral Right Angle Prism

S. Viridi[1] and Hendro[2]

[1]Nuclear Physics and Biophysics, Institut Teknologi Bandung, Bandung 40132, Indonesia
[2]Theoretical High Energy Physics and Instrumentation, Institut Teknologi Bandung, Bandung 40132, Indonesia
[1]dudung@fi.itb.ac.id, [2]hendro@fi.itb.ac.id

*Abstract*

*Trace of ray deviated by a prism, which is common in a TIR (total internal reflection) measurement system, is sometimes difficult to manage, especially if the prism is an equilateral right angle prism (ERAP). The point where the ray is reflected inside the right-angle prism is also changed as the angle of incident ray changed. In an ATR (attenuated total reflectance) measurement system, range of this point determines size of sample. Using JavaScript and HTML5 model and visualization of ray tracing deviated by an ERAP is perform and reported in this work. Some data are obtained from this visualization and an empirical relations between angle of incident ray source* $\theta_S$*, angle of ray detector hand* $\theta'_D$*, and angle of ray detector* $\theta_D$ *are presented for radial position of ray source* $R_S = 25 \text{ cm}$*, radial position of ray detector* $R_D = 20 \text{ cm}$*, height of right-angle prism* $t = 15 \text{ cm}$*, and refractive index of the prism* $n = 1.5$*.*

## Introduction

Various forms of prism have been used in diverse optical measurement systems, such as quadrilateral prism in telescope [1, 2], trapezoid prism in measuring thin film properties [3, 4], rectangular prism in phase-velocity measurements [5], penta prism to obtain high accuracy right angle deviation [6], hollow equilateral prism for fluid refractive index measurement [7], common right angle prism for apex angle measurement [8], and equilateral right angle prism (ERAP) in angle measurement [9] and object profile measurement [10]. The last form of prism or ERAP can also be used for thickness measurement of ultrathin films [11], wide fields of view scan [12], and polarization maintaining [13]. It is reported that the use of ERAP in ATR (attenuated total reflectance) measurement system delivers difficulty in adjusting position of ray detector since it must be adjusted every time position of ray source is changed, not only the position but also its orientation [14]. In this work the relation between ray source position, ray detector position, and ray detector orientation will be investigated. Model and visualization are built using JavaScript and HTML5.

**Theory**

An ERAP is positioned that its height $t$ toward $x$ axis as illustrated in Figure 1. This ERAP will be always in a fixed position and also will not be rotated. Three points of ERAP are defined as $(x_{\mathrm{PS}}, y_{\mathrm{PS}})$, $(x_{\mathrm{PE}}, y_{\mathrm{PE}})$, and $(x_{\mathrm{PN}}, y_{\mathrm{PN}})$. Subscript $\mathrm{P}$ indicates prism, while $\mathrm{S}$, $\mathrm{E}$, and $\mathrm{N}$ stand for south, east, and north respectively.

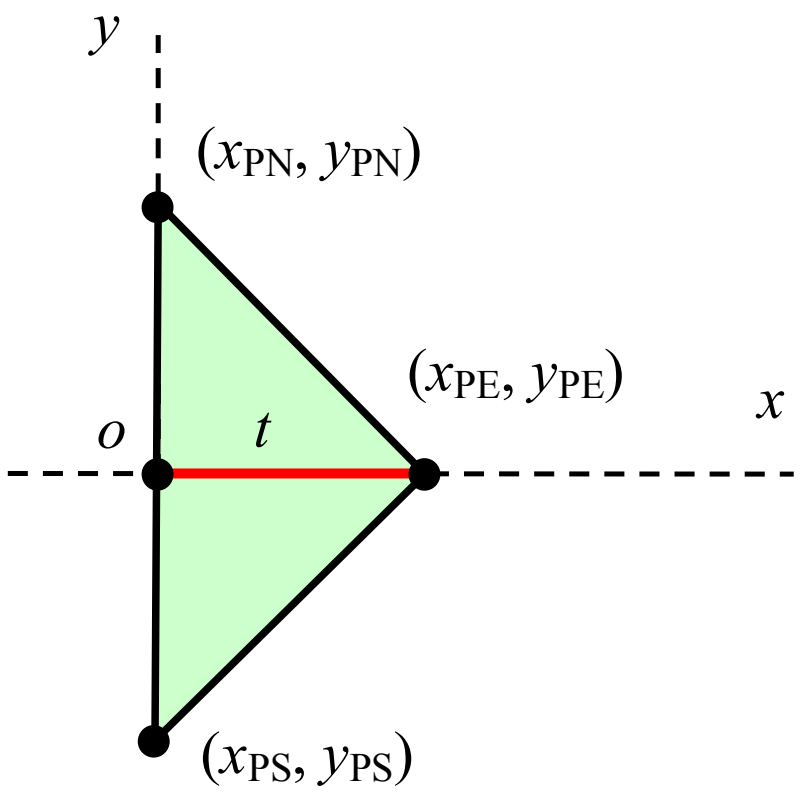

Figure 1. Configuration of ERAP with three points $(x_{\mathrm{PS}}, y_{\mathrm{PS}})$, $(x_{\mathrm{PE}}, y_{\mathrm{PE}})$, and $(x_{\mathrm{PN}}, y_{\mathrm{PN}})$ for the corners coordinates.

These points are not independent to each other. The relations between them are

$$x_{\mathrm{PN}} = x_{\mathrm{PS}}, \tag{1.a}$$

$$x_{\mathrm{PN}} - x_{\mathrm{PE}} = t, \tag{1.b}$$

$$y_{\mathrm{PN}} - y_{\mathrm{PE}} = t, \tag{1.c}$$

$$y_{\mathrm{PE}} - y_{\mathrm{PS}} = t. \tag{1.d}$$

And in order to simplify linear equation that will be defined, the origin of coordinate system is chosen as in Figure 1, where point $o$ is at $(0, 0)$.

Two refraction planes and one reflection plane are defined using linear equation. Each plane has its own linear equation. The first refraction plane labeled with subscript $\mathrm{SE}$ (southeast), which are defined using points $(x_{\mathrm{PS}}, y_{\mathrm{PS}})$ and $(x_{\mathrm{PE}}, y_{\mathrm{PE}})$, has linear equation

$$y_{\mathrm{SE}}(x) = \left( \frac{y_{\mathrm{PE}} - y_{\mathrm{PS}}}{x_{\mathrm{PE}} - x_{\mathrm{PS}}} \right) (x - x_{\mathrm{PS}}) + y_{\mathrm{PS}}. \tag{2}$$

The second refraction plane labeled with subscript NE (northeast), which are defined using points $(x_{\rm PN}, y_{\rm PN})$ and $(x_{\rm PE}, y_{\rm PE})$, has linear equation

$$y_{\rm NE}(x) = \left( \frac{y_{\rm PE} - y_{\rm PN}}{x_{\rm PE} - x_{\rm PN}} \right)(x - x_{\rm PN}) + y_{\rm PN} . \tag{3}$$

The only reflection plane labeled with subscript W (west), which are defined using points $(x_{\rm PN}, y_{\rm PN})$ and $(x_{\rm PS}, y_{\rm PS})$, has linear equation

$$x_{\rm W} = 0 , \tag{4}$$

since it is a vertical line.

A ray emitted from a source located at $(x_{\rm S}, y_{\rm S})$ towards coordinate origin $(0,0)$ propagates a long a linear equation

$$y_{\rm SS}(x) = \left( \frac{y_{\rm S}}{x_{\rm S}} \right) x . \tag{5}$$

Point where the ray hit the prism is intersection of Equations (3) and (5) if $y_{\rm S} > 0$ and intersection of Equations (2) and (4) if $y_{\rm S} < 0$. The point is defined as $(x_0, y_0)$.

From point $(x_0, y_0)$ a ray will be refracted according to Snell's law of refraction [15]

$$\frac{\sin \theta_{0i}}{\sin \theta_{0f}} = \frac{n_p}{1} , \tag{6}$$

where $\theta_{0i}$, $\theta_{0f}$, and $n_p$ stand for incident angle at point $o$, refraction angle at point $o$, and prism refractive index. Number 1 is used as denominator since the medium outside the prism is air. Normal line of the first refraction plane can be obtained from

$$\theta_{\rm NE} = {\rm atan} \left( \frac{y_{\rm PE} - y_{\rm PN}}{x_{\rm PE} - x_{\rm PN}} \right) + \frac{\pi}{2} , \tag{7}$$

while the angle of incident line is

$$\theta_{\rm SS} = {\rm atan} \left( \frac{y_{\rm S}}{x_{\rm S}} \right) . \tag{8}$$

From Equations (7) and (8) incident angle on the first refracted plane can be obtained

$$\theta_{0i} = \theta_{\rm NE} - \theta_{\rm SS} . \tag{9}$$

Expression in Equation (7) is an example if ray source is from first quadrant or $y_S > 0$. If ray source from fourth quadrant or $y_S < 0$, then Equation (7) must use Equation (2) instead of Equation (3).

From point $(x_0, y_0)$, the ray must propagate along a line with linear equation

$$y_{S0}(x) = \tan(\theta_{NE} + \theta_{0f})x + [y_0 - \tan(\theta_{NE} + \theta_{0f})x_0] \ . \tag{10}$$

This line intersects reflection plane defined by Equation (4), which gives point $(x_1, y_1)$.

On the reflection plane incident angle is obtained from

$$\theta_{1i} = 0 - (\theta_{NE} + \theta_{0f}), \tag{11}$$

and the reflection angle will be the same

$$\theta_{1f} = \theta_{1i}, \tag{12}$$

according to Snell's law of reflection [15]. Value $0$ in Equation (12) is obtained from the direction of normal line from the reflection plane.

Similar to Equation (10), the reflected ray must propagate from $(x_1, y_1)$ along a line described by linear equation

$$y_{S1}(x) = \tan(0 + \theta_{1f})x + [y_1 - \tan(0 + \theta_{1f})x_1] \ . \tag{13}$$

The next step is to find the intersect point between Equation (13) and the second refraction plane, which is defined by Equation (2). The point is labeled as $(x_2, y_2)$. Normal of this refraction plane is obtained similar to Equation (7), which gives

$$\theta_{SE} = \operatorname{atan}\left(\frac{y_{PE} - y_{PS}}{x_{PE} - x_{PS}}\right) - \frac{\pi}{2}. \tag{14}$$

At this point, the ray is refracted from denser medium to lesser medium, so that the Snell's law of refraction becomes

$$\frac{\sin\theta_{2i}}{\sin\theta_{2f}} = \frac{1}{n_p}, \tag{15}$$

with

$$\theta_{2i} = 0 - (0 + \theta_{1f}). \tag{16}$$

From point $(x_2, y_2)$, the ray suppose to hit the detector. The ray propagates along a line defined by

$$y_{S2}(x) = \tan(0+\theta_{2f})x + [y_2 - \tan(0+\theta_{2f})x_2] \ . \tag{17}$$

Illustration of points $(x_S, y_S)$, $(x_0, y_0)$, $(x_1, y_1)$, $(x_2, y_2)$, $y_{SS}(x)$, $y_{S0}(x)$, $y_{S1}(x)$, and $y_{S2}(x)$ is given in Figure 2.

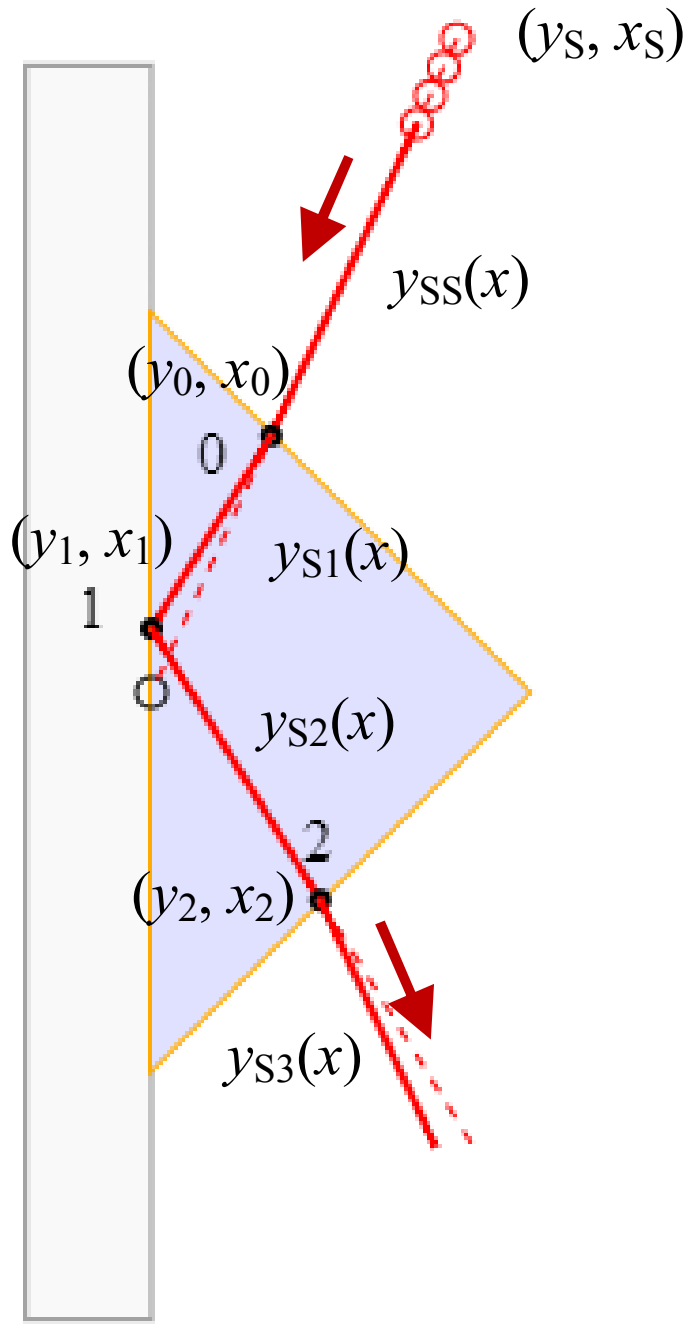

Figure 2. Tracing of the ray of an ERAP with source located in first quadrant ($y_S > 0$) and position of points $(x_S, y_S)$, $(x_0, y_0)$, $(x_1, y_1)$, and $(x_2, y_2)$ and the liniear equations $y_{SS}(x)$, $y_{S0}(x)$, $y_{S1}(x)$, and $y_{S2}(x)$.

**Model and visualization**

Model of ray tracing is represented by points which are defined by Equations (1) – (17) applying only Snell's law of refraction and reflection and linear equation. After obtaining $(x_S, y_S)$, $(x_0, y_0)$, $(x_1, y_1)$, $(x_2, y_2)$, and also $(x_3, y_3)$ the trace of ray can be visualized. The last points $(x_3, y_3)$ is the outmost position where prism deviated ray should be drawn.

Visualization of the ray tracing is constructed using HTML5 and its JavaScript implementation. Google Chrome version 29.0.1547.66 m is used as internet browser to show the model and visualization result. Three files are generated in this work, which are `atr.html`, `style.css`, and `script.js`. All files can be found in here [16].

Details of HTML5, JavaScript, and CSS can found in various resources in internet, e.g. W3Schools [17].

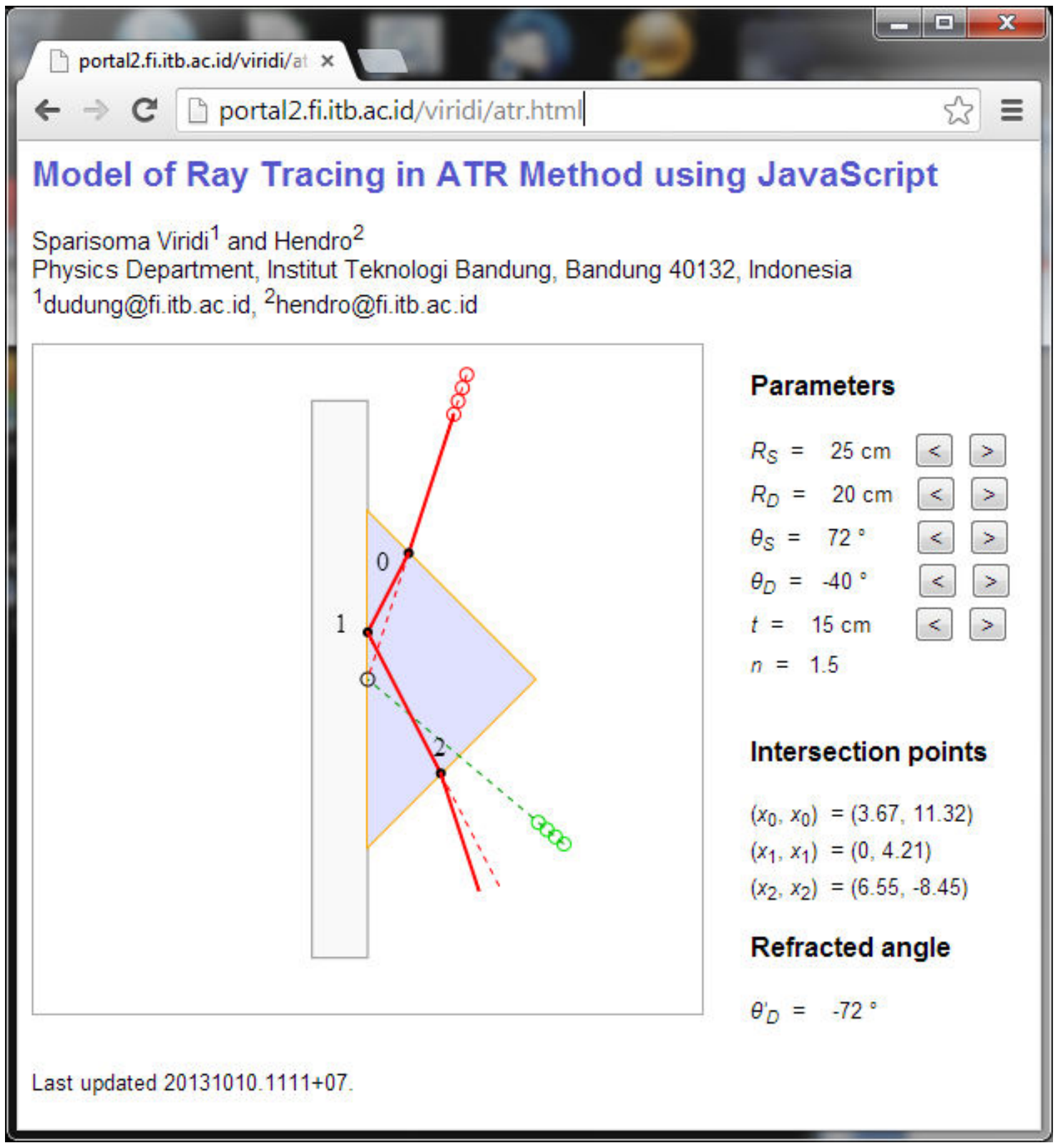

Figure 3. Visualization result using HTML5, JavaScript, and CSS [16].

## Results and discussion

Some parameters are used in the model and visualization, whose symbol and its meaning is given in Table 1.

Tabel 1. Parameters used in the model and visualization.

| Symbol | Unit | Meaning |
|---|---|---|
| $R_S$ | cm | radial position of ray source |
| $\theta_S$ | ° | angular position of ray source* |
| $R_D$ | cm | radial position of ray detector* |
| $\theta_D$ | ° | angular position of ray detector |
| $\theta'_D$ | ° | orientation of detector* |
| $t$ | cm | thickness of ERAP |
| $n$ | - | prism refractive index |

*Angle is measured from $x$ axis in counter clockwise direction.

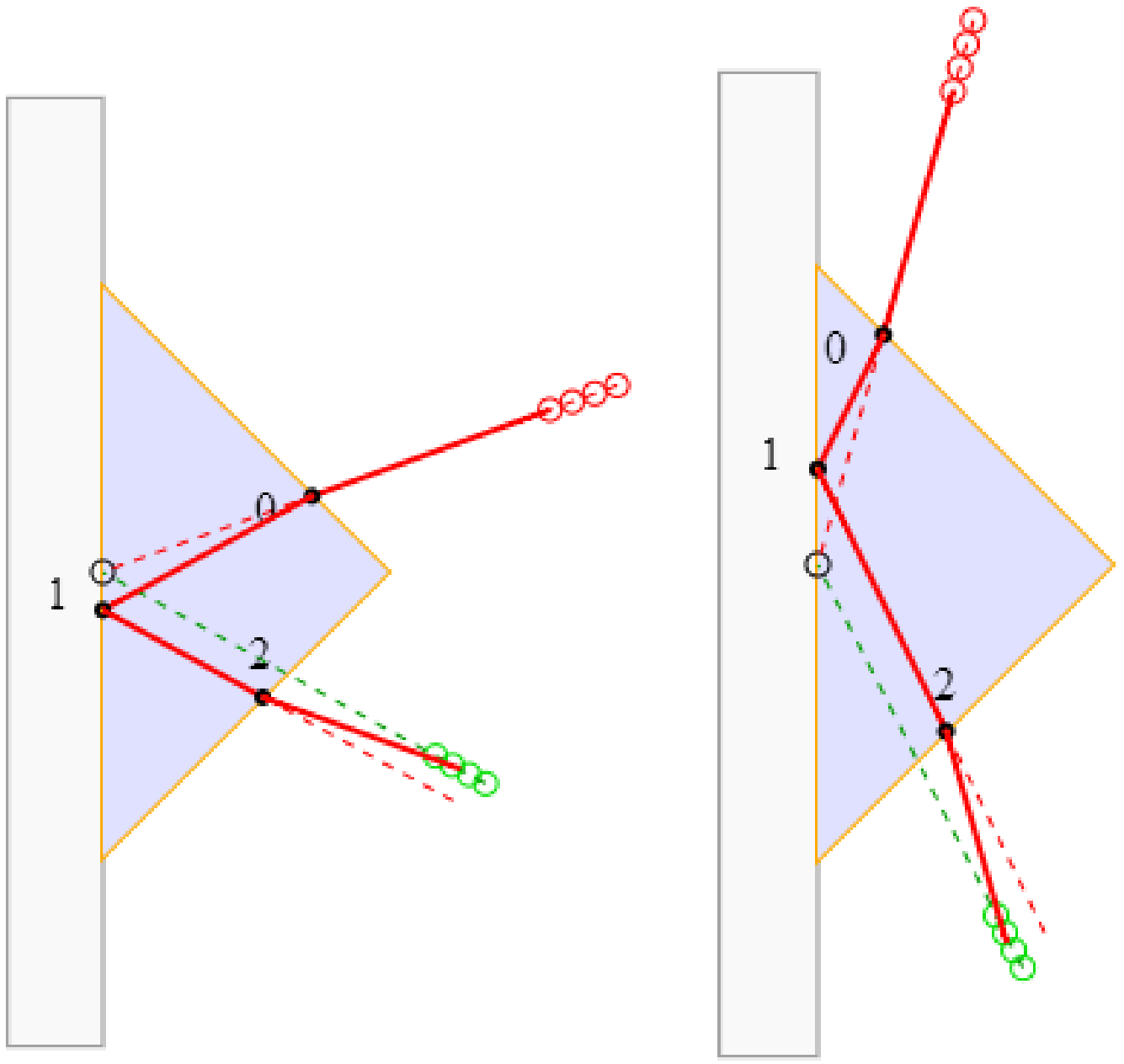

Figure 4. Results of ray tracing with $R_S$ = 25 cm, $R_D$ = 20 cm, $t$ = 15 cm, and $n$ = 1.5 for: $\theta_S$ = 20 °, $\theta_D$ = -29 °, $\theta'_D$ = -20 ° (left) and $\theta_S$ = 74 °, $\theta_D$ = -63 °, $\theta'_D$ = -74 ° (right).

Result of the visualization can be seen in Figure 3, where it can also be access remotely since it is build using HTML and viewed with an internet browser. For two different parameters, examples of ray tracing are given in Figure 4, where left part if for small angular position of ray source, while right part is for large angular position of it.

Relations between $\theta_S$, $\theta_D$, $\theta'_D$, and $y_1$ are given in Figure 5 accompanied with empirical equations. Relation between $\theta_S$ and $\theta'_D$ is not given since both have the same value as in a parallel prism. From Figure 5.(a) the relation between $\theta_D$ and $\theta_D$ is almost linear but still dependent detector radius $R_D$. Information of $\theta_S$ - $\theta_D$ value in Figure 5.(b) might be useful in designing automatic positioning of ray detector as function of ray source position. Orientation of detector $\theta'_D$ on detector position $\theta_D$ as given in Figure 5.(c) shows that the use of ERAP is not easy as the use of semicircular prism. In the case of semicircular prism it always holds that $\theta'_D = \theta_D$, which makes the setup very easy. The last parameter which is also important is $y_1$, where its relation with $\theta_S$ is given in Figure 5.(d). This parameter limits the size of sample which is used, e.g. in ATR method. If the sample is smaller than value of $y_1$ in some certain $\theta_S$, then at that angle there will be no valid signal detected in detector. Observer should be aware of this limitation before the experiment is performed.

Further plan is to generate more general data that also shows the dependence to prism refractive index and detector radius. It also would be better when the relation of $\theta_S$, $\theta_D$, $\theta'_D$, and $y_1$ can be obtained analytically.

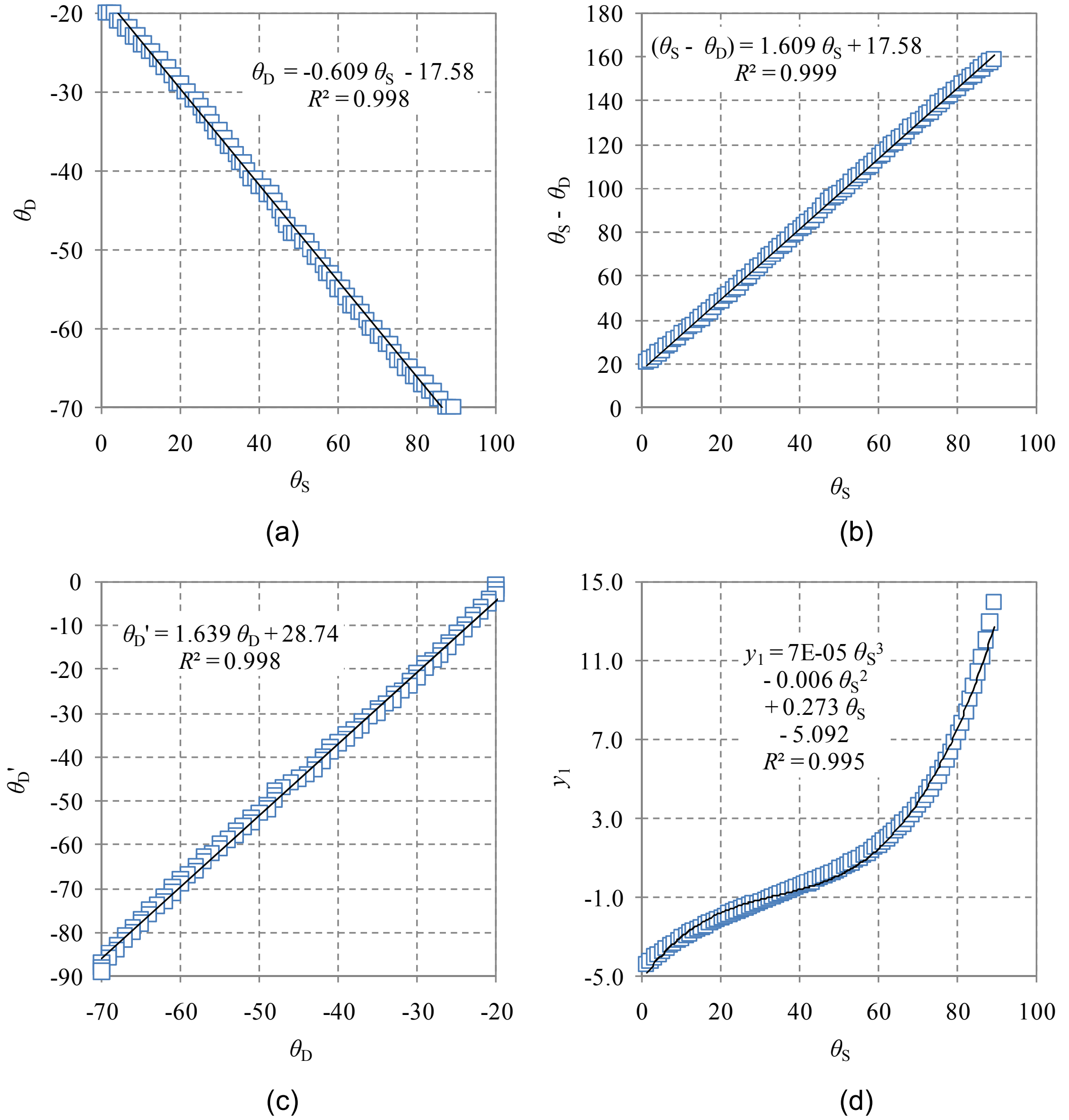

Figure 5. Relations between $\theta_S$, $\theta_D$, $\theta'_D$, and $y_1$.

## Conclusion

In this work model and visualization of ray tracing deviated by an ERAP (equilateral right angle prism) has been performed using HTML5 and JavaScript, enhanced with CSS to make the view more user friendly. As the results four empirical relation between $\theta_S$, $\theta_D$, $\theta'_D$, and $y_1$ has been obtained. Value of $R^2$ of these relations are more than 0.995.

## Acknowledgements

This work is supported partially by Riset Inovasi KK (RIK) ITB in year 2013 with contract number 248/I.1.C01/PL/2013.